# Some aspects of the field evaporation behaviour of GaSb

B. Gault[1,2]*, D.W. Saxey[1,3], G.D.W. Smith[1], M. Müller[1]

[1]*Department of Materials, University of Oxford, Parks Road, Oxford, OX13PH, UK*

[2] *now at The Australian Centre for Microscopy and Microanalysis, Madsen Building F09, The University of Sydney, NSW 2006, Australia*

[3] *now at School of Physics, The University of Western Australia, CRAWLEY WA 6009, Australia*

*corresponding author: baptiste.gault@sydney.edu.au

**Abstract**

Molecular or cluster ions are often observed in the atom probe microanalysis of III-V compound semiconductors. Here, in-depth data analysis of a series of experiments on GaSb reveals strong variations in the mass spectrum, cluster ion appearance and multiplicity of the detector-events with respect to the effective electric field at the specimen surface. These variations are discussed in comparison with Al 6XXX series alloys and pure W and it is proposed that they may originate from field-dissociation of molecular ions, which might contribute to compositional inaccuracies.

# 1 Introduction

Major breakthroughs in the design of atom probe tomography (APT) through the past decade have enabled an unprecedented broadening of the field-of-application of the technique [1-2]. In particular the implementation of laser pulsing capabilities [3-8], has made possible routine analysis of semiconducting materials. APT relies on the time-controlled removal of surface atoms in the form of ions, from a needle-shaped specimen, induced by a very intense electric field in a process known as field evaporation [9-10]. Field evaporation is a thermally activated process, and in pulsed-laser APT the field evaporation is triggered by the transient, spatially and temporally confined increase in temperature induced by absorption of the laser light by the specimen [11]. The surface is subsequently quenched as the heat is transferred along the specimen axis [12].

However the field evaporation behaviour of semiconductors in the atom probe can be significantly different from that of metals, as demonstrated by early investigations of semiconductors, where molecular or cluster ions were observed [13-15]. More recently, compound semiconductors were reported to exhibit similar behaviour [16-18]. Although the mechanism of formation of these cluster ions is still unclear, their presence during atom probe analyses seems to depend strongly on the experimental conditions. Cerezo *et al.* [19] proposed to use their relative abundances to estimate the specimen temperature in a methodology similar to that proposed by Kellogg based on charge-state ratios [20]. Furthermore, atom probe analyses of poorly conducting materials exhibits a high number of multiple events, where a single laser pulse induces detection of significantly more than a single ion per pulse [21]. Here, we report on the analysis of gallium antimonide across a wide range of experimental conditions, and investigate the field evaporation behaviour of this material in contrast to that of pure W and an Al-6XXX series alloy in order to optimise the accuracy of atom probe microanalysis of compound semiconductors.

## 2 Experimental

The material under investigation is a 0.75 µm thick un-doped layer of gallium antimonide (GaSb) grown on a (100) GaSb substrate via molecular beam epitaxy (MBE). The layer growth was carried out at a temperature of 500 °C with a growth rate of 0.74 ML/s. The MBE-grown GaSb layer is assumed to be stoichiometric down to a few $10^{17}$ cm$^{-3}$.

Specimens for atom probe analysis were prepared via *in-situ* lift-out in a Zeiss NVision 40 scanning electron microscope equipped with a focused ion beam and a Kleindiek micromanipulator. The procedure followed for the preparation is similar to the one proposed by Thompson *et al.* [22]. Relatively low (40-150pA) ion currents were used as the material appeared to be sensitive to the beam. To minimise ion-beam damage, a protective layer comprising a ~ 130 nm thick ex-situ sputtered gold layer and a 200 nm thick ion-beam deposited tungsten-carbon layer was deposited on the wafer surface.

Experiments on metallic materials were carried out under comparable conditions on the same instrument, with specimens of pure W, prepared from ultra-high purity (99.99%) W wire by electrochemical polishing in a solution of NaOH with ~ 5V, and of an Al-based alloy from the 6XXX series (containing 0.56 at% Mg and 0.96 at% Si), prepared using a solution of 10% Perchloric acid in Butoxyethanol at 2-20V.

Atom probe experiments were performed on a Cameca LEAP™ 3000X HR in laser-assisted mode. Pulses were generated at 100kHz by a Nd:YVO solid-state laser (duration 10 ps, wavelength 532nm). The beam was focused down to a diameter < 10μm. The laser pulse energy was varied in the range 0.01-0.1 nJ using a base temperature of ~ 43K at an average detection rate of 0.01 ion/pulse. The detector efficiency was approximately 37%. Experiments with constant laser energy and varying detection rates (0.002, 0.01, 0.02 and 0.03 ions/pulse) or specimen base temperatures (43, 88, 118 and 149 K) were also performed. Each dataset contained 2-3 million ions. Further, the variable parameters were not progressively increased, but changed in a random order to minimise biases. Data reconstruction was performed using IVAS™, commercial software, and multiple detector hit analyses were made possible thanks to the provision of access to Imago Root™.

## 3 Methods

### 3.1 Mass spectrum analysis

The mass spectra shown in Figure 1 exhibit a large number of cluster ions, mostly containing Sb. As Sb has two isotopes, mass peak deconvolution was employed to assess the relative quantities of $Sb_n^+$ and $Sb_{2n}^{2+}$ within overlapped peaks, as described in [23]. A manual background correction was performed by averaging the background counts in ranges of equivalent size within the mass spectrum on each side of each peak to adjust the measured quantities of each species. All the compositions given in this paper were corrected to account for both the peak overlaps and the local level of background.

### 3.2 Estimation of the effective evaporation field

Due to changes in the laser energy or specimen geometry amongst other parameters, the actual electric field conditions differ at the specimen apex in each data set. Variations of the electric field can be traced by a change in the charge state of the ions, which in turn, can be used to estimate the electric field, using post-ionisation theory [24-25], and to assess the experimental conditions [26]. Here, Kingham-like curves for Ga were computed using the equations from ref. [25] and used to estimate the electric field for each set of experimental conditions (laser energy, base temperature, detection rate). This value represents the effective evaporation field $F_{eff}$ required to induce field evaporation under these conditions.

### 3.3 Analysis of multiple events

For each individual pulse, the number of ions reconstructed from the data of the delay-line detector is recorded. Access to this information enables in-depth investigation of the field evaporation behaviour [21, 27-28]. The histogram of the number of ions contained in each event, referred to as multiplicity, is plotted in Figure 3. Ions can also be filtered based on their multiplicity, allowing investigation of spatial or temporal correlations between ionic species [27].

A methodology was developed to investigate correlations between ionic species within multiple events. A symmetric $n \times n$ matrix, or table, $p_{ij}$, is generated by counting coincidences of ion types '$i$' with '$j$' within a multiple event. '$n$' is the number of ion types considered; e.g. the more abundant species within the mass spectra. Other mass values are considered as background and ignored. A contingency table-type analysis applied to the matrix values reveals which ion-pairs appear more within the data than expected from independent correlated evaporation events [27]. Here, in contrast to the application of contingency tables to assess spatial correlations within atom probe data [29-31], each element of the table is compared with its own expectation value to determine the strength of correlation between the corresponding $i,j$ ions within multiple events. More details on this analysis method can be found in ref [32].

## 4 Results and discussion

*4.1 Mass spectra*

Figure 1 shows mass spectra obtained for different $F_{eff}$, where relevant peaks are labelled, $F_{eff}$ and the proportion of multiple events are specified. Heavy cluster ions, up to $Sb_5$, and mixed GaSb ions progressively appear as $F_{eff}$ decreases. Conversely, at higher $F_{eff}$, $Sb_1$ and $Sb_2$ dominate. As $F_{eff}$ is reduced the level of background significantly decreases and the tails of peaks becomes slightly steeper. The increase from ~6 to ~ 45% of multiple events between low and high field conditions represents an unusual proportion that complicates data interpretation.

Ga ions are almost never detected as part of molecular ions. Determination of $F_{eff}$ based on the charge-state ratio relies on the physics of electric-field-induced post-ionisation, and therefore Ga is a suitable candidate for that purpose. Detection of mixed Ga-Sb cluster ions might induce some inaccuracies in the estimation of $F_{eff}$ in the lower range of electric field estimations.

*4.2. Chemical accuracy*

Regarding optimisation of the conditions for atom probe microanalysis of GaSb materials, the aim is to find conditions where the expected III/V (Ga/Sb) ratio of 1 is obtained. The III/V ratio is plotted in Figure 2 as a function of $F_{eff}$. An excess of the Group III element (Ga) is observed across the range of $F_{eff}$. This could be regarded as an excess of Ga within the material or the specimen. Typically, MBE fabrication yields an accurate stoichiometry, and this hypothesis seems rather unlikely. The specimen preparation involves a focused Ga ion beam, and ion implantation is a known problem [22, 33]. A low energy cleaning procedure removes the highly implanted region from the specimen, thereby reducing bias in the analysis. If the excess of Ga was due to such implantation, it should progressively decrease as the specimen is analysed, in contrast to what is actually observed. Variations seem intrinsically linked to the experimental conditions. Poor control of the laser illumination conditions was proposed to lead to overheating of the specimen, inducing migration of Ga along the shank of the specimen yielding an excess of Ga [34]. Here, the illuminated region is minimised and no evidence of pronounced surface migration was observed in the form of density variations within the field desorption maps [35], indicating that this process is unlikely to affect the data.

An excess of Ga could relate to preferential field evaporation of Sb or antimony clusters, which is supported by the decrease in the background as $F_{eff}$ decreases. The formation of heavy ions could also partly explain the loss of Sb at low $F_{eff}$. The detector only registers impacts arriving within a time window after the emission of a pulse. Sb-containing heavy ion clusters could be specifically lost due to too long times-of-flight. Further, at 19.75 Vnm$^{-1}$, ~ 30% of Ga ions and ~ 45% of all Sb ions are detected as part of multiple events, and this peaks at 60% for Sb$_1$. These quantities drop to only ~ 5% of Ga and ~10% of Sb at 16.25Vnm$^{-1}$. On multiple events, ion pile-up, or intrinsic limitations of the delay-line detector system due to an overload of signals to treat, can lead to a loss of some ions. At high $F_{eff}$ Sb might be specifically lost due to detector limitations, while at low $F_{eff}$ Sb might be lost due to non-registration of heavy ion impact. This explains the presence of optimal analysis conditions where a balance is found between the amount of multiple events and the formation of large clusters.

*4.3. Multiple events*

Normalised histograms of the event multiplicity are displayed in Figure 3 for GaSb, with three values of $F_{eff}$, pure W and Al-6XXX. For metallic materials, multiple events are generally attributed to correlated field evaporation, a process whereby an enhanced instability of surface atoms is observed after one of their nearest neighbours has been field evaporated [27]. Atoms in the vicinity of atoms that were recently evaporated are thus are more prone to field evaporate either on the same pulse or on the next few pulses. It was demonstrated that as the electric field is increased, these events take place more often [27].

Here, for a similar increase in $F_{eff}$, the proportion of multiple events in GaSb is observed to increase 3 to 4 times more than in the Al6XXX alloy. Such a difference in behaviour might find an origin in the electronic properties of semiconductors that significantly differ from those of metallic materials, in particular the mean-free path of the electrons. To accommodate changes in the electric field, electron displacements over long distances could occur, taking relatively long times and hence enhancing the probability of correlated field evaporation and increasing the range of distance where correlated evaporation can occur. Another possible explanation of the very high rate of multiple events is field-dissociation: as cluster ions travel through the high electric field, they can experience post-ionisation or field-dissociation or both [36]. Field dissociation of ions leads to complete or partial fragmentation of these clusters and hence generates several atomic or molecular ions that would lead to the detection of multiple events.

*4.4. Cluster ions abundances*

The relative frequency of the different cluster ions encountered in experiments with varying laser pulse energy is plotted as a function of $F_{eff}$ in Figure 4(a), while in Figure 4(b), the charge-state ratio of $Sb_3$, a particularly stable cluster ion, is plotted as a function of $F_{eff}$. For experiments at constant laser pulse energy with varying average ion detection rate, the relative frequency of the different cluster ions is plotted vs. $F_{eff}$. An increase in detection rate arises from an increase in the DC voltage applied to the specimen and thus in $F_{eff}$. Results are in good agreement with those shown in Figure 4(a).

The abundances of cluster ions reflect the relative stability of the clusters with respect to fragmentation [37]. Together with the charge state of stable clusters, it can be used to efficiently trace the electric field conditions (Figure 4(a)-(b)). These networks of curves are the counterpart for cluster ions of those from the post-ionisation theory. $Sb_1$ dominates under high field strengths (~95 % of all Sb), but under the lowest field strength $Sb_3$ dominates. $Sb_3$ contains one of the *magic numbers* of atoms and is amongst the most stable cluster ions due to favourable configurations of the skeletal electrons [38-39]. Clusters of 5 atoms are also observed under low electric field conditions

In the experimental results presented in Figure 4(c), only $F_{eff}$ is changed. The temperature reached by the specimen can be assumed to be constant and so the type and amount of cluster ions formed should be relatively constant. Hence, $F_{eff}$ significantly contributes to the variations in the relative abundances of cluster ions detected. For example, smaller clusters are detected as $F_{eff}$ was increased. These results are suggestive of field dissociation, as $F_{eff}$ has a significant impact on the detection of cluster ions.

*4.5. Spatial correlations*

Atoms arriving on events with a multiplicity of 2 were first selected to investigate potential field-dissociation events. The distance between the impacts of the two ions on the detector was computed, rather than the distance within the three-dimensional reconstruction, in order to avoid potential inaccuracies in the estimation of the magnification and allow for direct comparison. Normalised histograms of these distances are plotted in Figure 5 for the analysis of W and Al-6XXX, as well as for GaSb with three different $F_{eff}$ values. For the two metallic materials, a strong peak is observed at small distances, similarly to previous observations [27, 40]. For GaSb, a peak appears at short distance and it tends to broaden as $F_{eff}$ increases. For $F_{eff}$ = 19.75 Vnm$^{-1}$, species specific pair distributions (Ga-Ga, Ga-Sb and Sb-Sb) exhibit significantly different behaviours. The Ga-Ga distribution forms a sharp peak at small distances and the Ga-Sb peak is comparable to the peak observed at lower $F_{eff}$. Finally, the peak for Sb-Sb pairs appears broader shifted towards larger distances, with a shape that suggest the overlap of two peaks. The difference in behaviour between Ga-Ga or Sb-Ga pairs and Sb-Sb pairs suggests that the former two could be linked to correlated evaporation, while the latter would be due to field dissociation, as ions generated via dissociation can have an excess of kinetic energy that might force them to deviate from their expected trajectories.

As introduced in ref. [27], correlated field evaporation not only affects ions detected on the same pulse, but also on successive pulses. Distances between ions arriving on two consecutive pulses were plotted in Figure 6 in red, as well as those coming on multiple events in blue. To reduce fluctuations due to limited statistics, the red curve was smoothed over 10 bins and it was normalised to the small distance shoulder of the blue curve to enable direct comparison. Both distributions were oversampled using a piecewise cubic spline interpolation method to enable estimation of their ratio (shown inset), which exhibits a peak for distances ~ 1.5 mm, highlighting the difference between the two distributions. These observations suggest that correlated field evaporation and field dissociation are both occurring simultaneously.

*4.6. Chemical correlations*

In searching for preferred correlations between ion species, a statistical analysis was performed on ion-pairs arriving within the same multiple event, for both low and high field conditions. The resulting correlation tables for GaSb are shown in Figure 7, where strong correlations are shown in red, while anti-correlations are coloured in green. Quantitatively the degree of correlation is indicated by the parameter

$$d_{ij} = \frac{p_{ij} - e_{ij}}{\sqrt{e_{ij}}},$$

where $p_{ij}$ is the observed number of *i-j* ion-pair co-incidences, and $e_{ij}$ is the expected number of co-incidences, calculated in the usual way from the row and column totals in the contingency tables [41]. Entries corresponding to low $p_{ij}$ or $e_{ij}$ values ($< 5$) are not considered to be statistically significant and are set to zero. Self-correlated events, on the table diagonal, are negatively biased by ion pile-up effects. This bias is removed by setting the diagonal elements ($p_{ii}$) equal to their expectation values so that their entries in the correlation table are close to zero. The abundance of each cluster ion type within multiple events is shown in Figure 7 to provide a measure of quantification.

For both high and low field conditions, the strongest correlation exists between the $Sb_2^+$ and $Sb^{2+}$ ions, which suggests a dissociation process, as correlated field evaporation between these particular species, to the exclusion of others, is otherwise difficult to explain. Assuming that these ions are products of a dissociation not involving any subsequent post-ionization or any other ion products, the parent cluster must be $Sb_3^{3+}$. $Sb_3$ is observed in the mass spectra in the 1+ and 2+ charge states, but not in a 3+ state. It may be that $Sb_3$ ions are commonly generated during field evaporation but the ions are unstable and dissociate into $Sb_2^+$ and $Sb^{2+}$. At higher $F_{eff}$, more $Sb_3$ ions are triply-ionized and then break apart, leading to fewer $Sb_3$ ions and more correlated $Sb_2^+/Sb^{2+}$, as observed here.

A few more general observations can be made here. At low $F_{eff}$, the Ga ions tend to arrive together with $Sb_n^{2+}$ cluster ions, while $Sb_1$ ions arrive with $Sb_n^{1+}$. This trend is still apparent for high $F_{eff}$ although there are less Sb clusters present and the Ga ions are more associated with single Sb ions than in low $F_{eff}$. Correlations at low $F_{eff}$ are more common when the parent cluster ion contains more Sb atoms or has a high charge state (>2+), which suggests that larger cluster ions with higher charge states tend to be less stable, in agreement with studies on carbon cluster ions [42]. Correlations between Ga and $Sb_n$ are common in both data sets, with n=3, 4, 5 for low $F_{eff}$ and n=1, 2, 3 for high $F_{eff}$.

It seems that the dissociation route might be mapped out by looking at correlations between species on multiple events. The origin of these cluster ions remains uncertain. They can be formed by bonding of monomers or multimers migrating over the surface. High-evaporation field adatoms are retained on the specimen surface, and can be subject to thermally activated surface migration and form multimers with other adatoms. Multimers are less tightly bonded to the surface than isolated adatoms which facilitates their field evaporation, as observed by field ion microscopy [43], and this is believed to lead to the formation of large clusters in liquid-metal ion sources [44]. This mechanism is assumed to be favoured by the higher temperatures reached in pulsed-laser APT. Hence heavier cluster ions would be expected to form as the temperature reached by the specimen during pulsed laser illumination increases. The relative abundances of cluster ions would thus be more influenced by the laser energy than by the electric field conditions. Another mechanism implies that several atoms covalently bonded within the material leave the surface as a whole. GaSb has a zinc-blende structure with a unit pattern consisting in a pyramid with a Ga surrounded by four Sb atoms and vice versa, which translates into each Sb atom being surrounded by a first shell of 4 Sb atoms [45]. The present investigation suggests that the formation of these cluster ions necessarily involves a combination of these two mechanisms, as both the temperature and electric field impact the observed abundance of cluster ions.

## 5 Summary and conclusion

We have shown that, in the atom probe microanalysis of GaSb, cluster ions are formed and can be dissociated in a mechanism that involves the intense electric field in the vicinity of the specimen surface. The appearance of these clusters is likely to be due to simultaneous field evaporation of groups of atoms originally bonded together within the lattice of the specimen, as no significant surface diffusion was observed. In-depth analysis of multiple events can help to cast light on the mechanisms leading to the dissociation of such clusters and the fragmentation route. Furthermore, it is worth noting that dissociative processes can also lead to the formation of neutral molecules [38] that would then be undetected in a reflectron-fitted atom probe microscope, such as the one used for these investigations. This mechanisim, combined with the very high proportion of multiple events, is likely to impact the chemical accuracy of the atom probe microanalysis of such materials.

**Acknowledgements**

Thanks to A. Cerezo for fruitful discussions. The authors are grateful to Teledyne Scientific Ltd., Thousand Oaks, CA, USA for provision of the materials used for this study. The authors are also grateful for scientific and technical input and support from the OPAL UK Atom-probe Facility at the University of Oxford – funded by the UK Engineering and Physical Sciences Research Council (EPSRC) under grant no. EP/077664/1. B. G. acknowledges the financial support of the European Commission via a FP7 Marie Curie IEF Action No. 237059. M. M. acknowledges the EPSRC Analytical Chemistry Trust Fund, Imago Scientific Instruments, the German National Academic Foundation, the "Hamburger Stiftung für Internationale Forschungs- und Studienvorhaben" and the German National Exchange Service (DAAD) for the generous scholarships provided.

**Figure Captions**

Figure 1: A comparison of mass spectra from GaSb at three different values of the average surface field, corresponding to different incident laser powers. The data set size and the evaporation rate are approximately equal in each case.

Figure 2: Ratio of detected group III and group V atoms as a function of the surface field. A spline was added as a guide for the eyes.

Figure 3: Histograms of detector event multiplicity for W, a 6xxx Al alloys, and for GaSb under three different conditions of the surface field.

Figure 4: The relative abundances of cluster ions and their dependence on the surface field. (a) $Sb_n$ cluster abundances, with field variation induced by laser power. (b) $Sb_3$ cluster charge-state ratio, with field variation induced by laser power. (c) $Sb_n$ cluster abundances, with field variation induced by evaporation rate.

Figure 5: Detector hit separations for ion-pairs in W, 6xxx Al alloy and GaSb. Several plots are shown for GaSb, corresponding to different surface fields and specific ion-pair combinations.

Figure 6: Detector hit separations for ion-pairs from GaSb (Ga-Ga, Sb-Ga, Sb-Sb): for ions evaporated from the same pulse, and for ions evaporated from successive pulses. Inset: Ratio of the distributions; same-pulse / successive-pulse.

Figure 7: Statistical analysis of ion-pair combinations arriving within the same multiple event. (a) High-field conditions (19.75 Vnm$^{-1}$). (b) Low field conditions (16.25Vnm$^{-1}$). Significant correlations (red) or anti-correlations (green) are indicated by strong colours. The numbers in the colour scale correspond to the *d* value defined in the text. The order of ions within each pair is ignored and so the tables are symmetric. Also plotted on the vertical axes are the number of ion-pairs of which each species is a member.

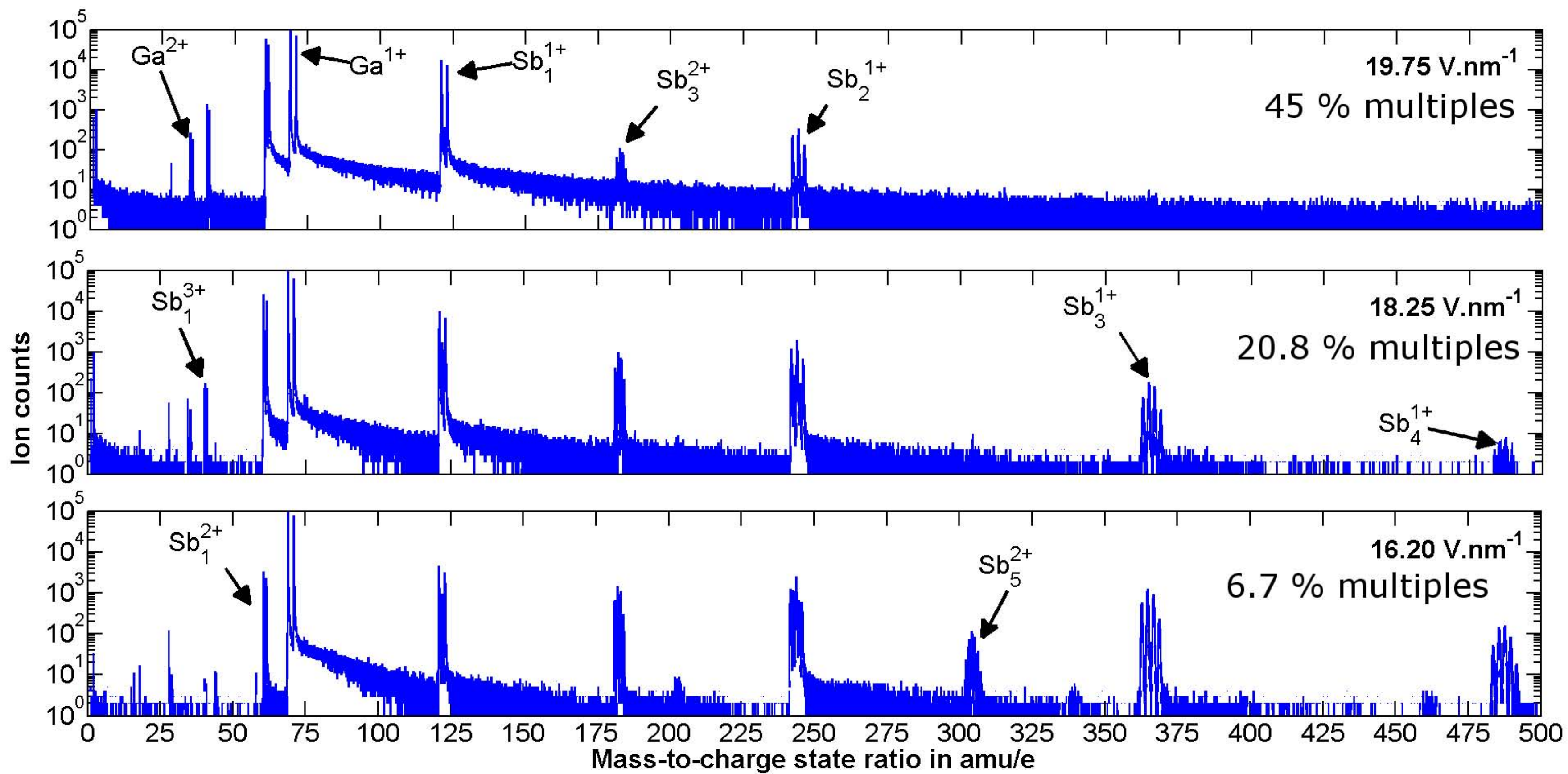

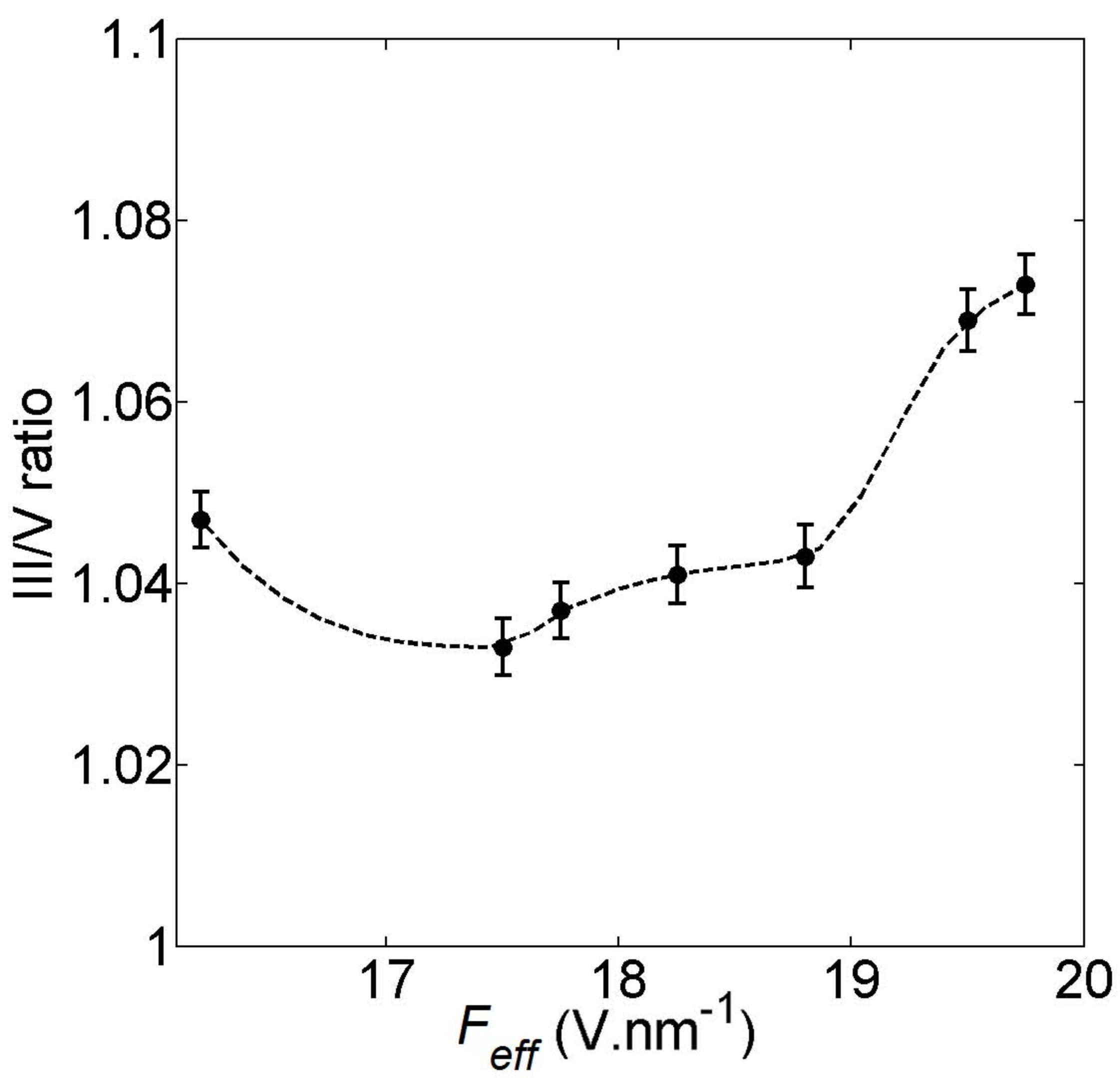

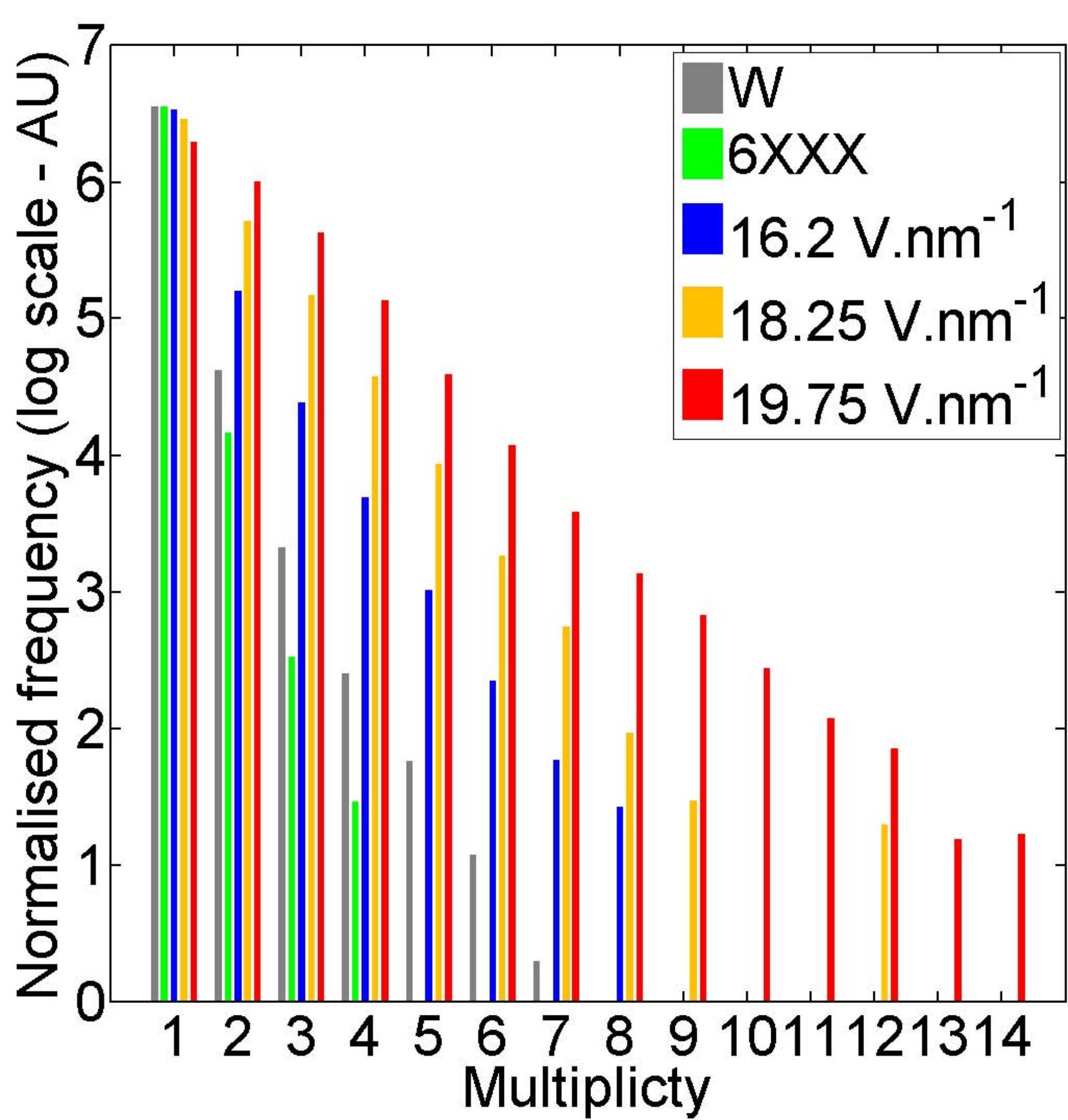

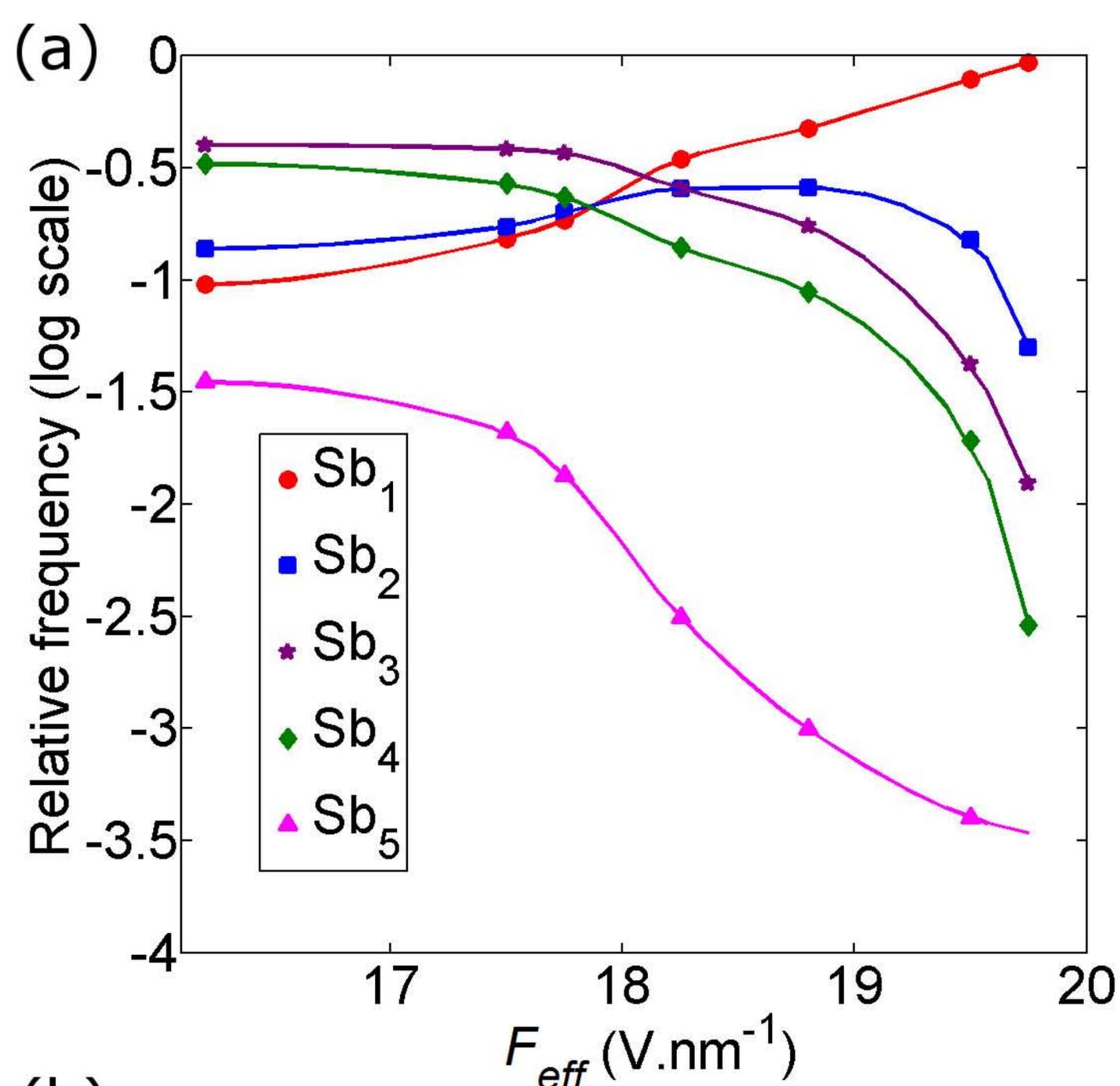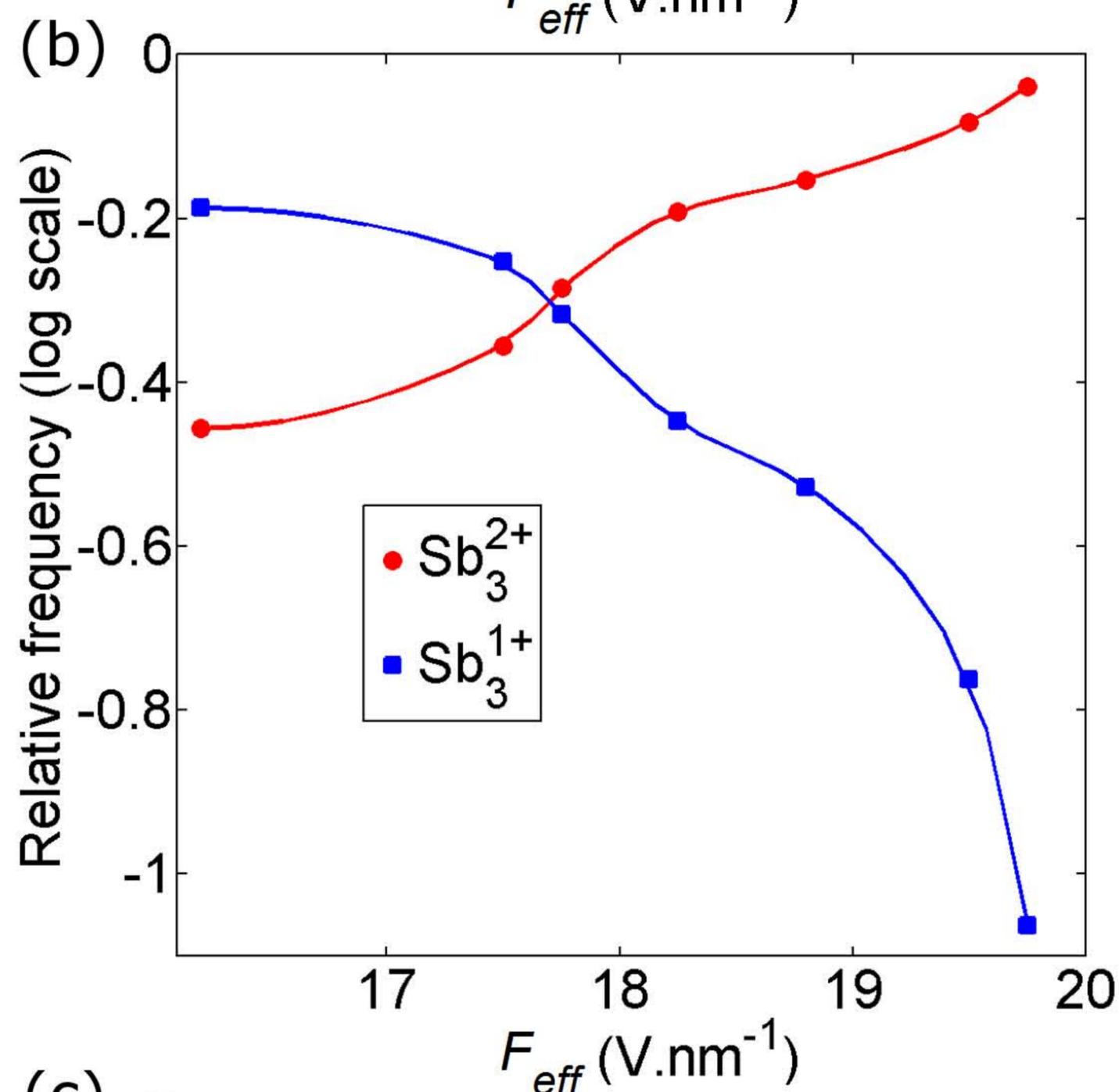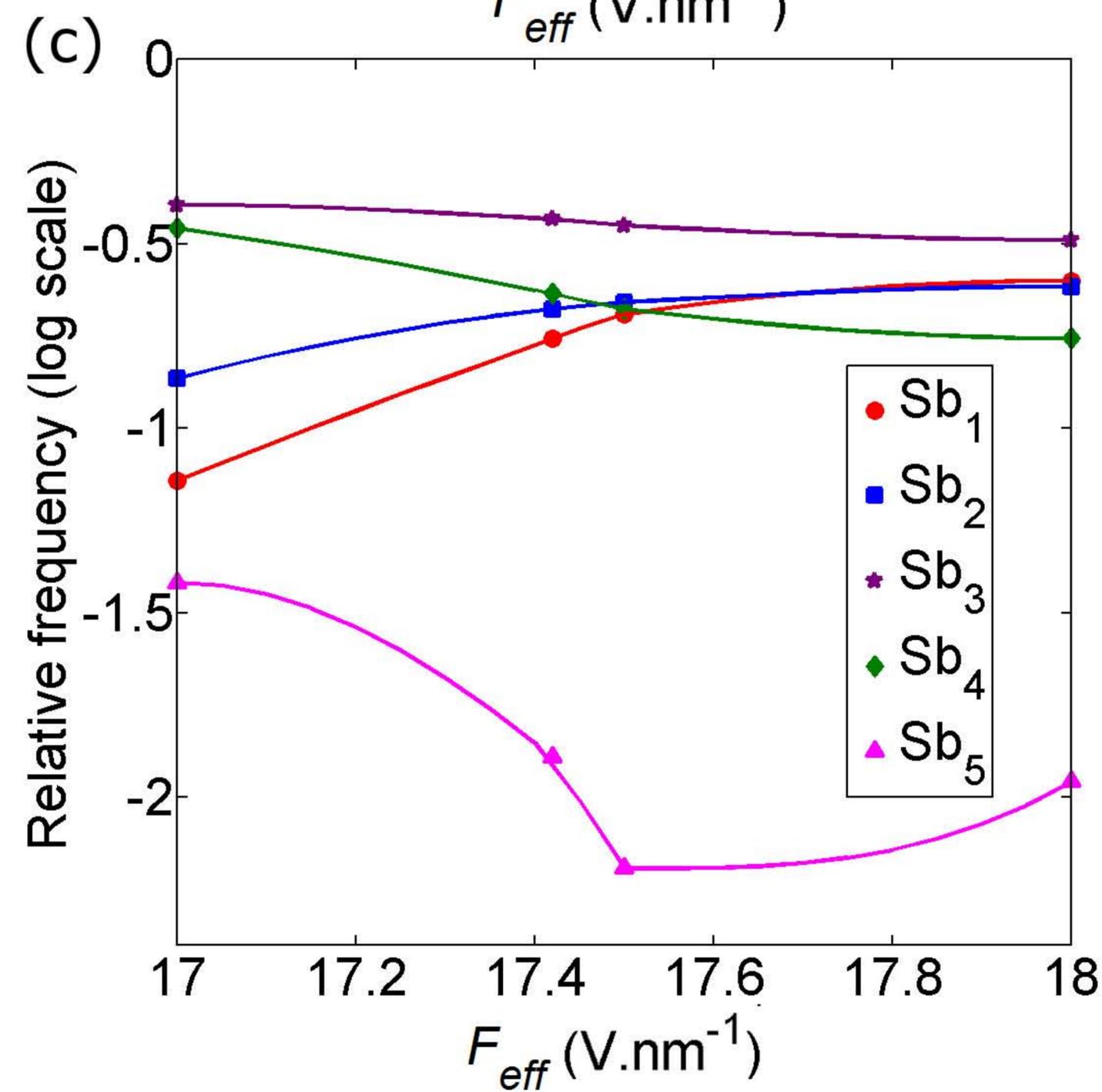

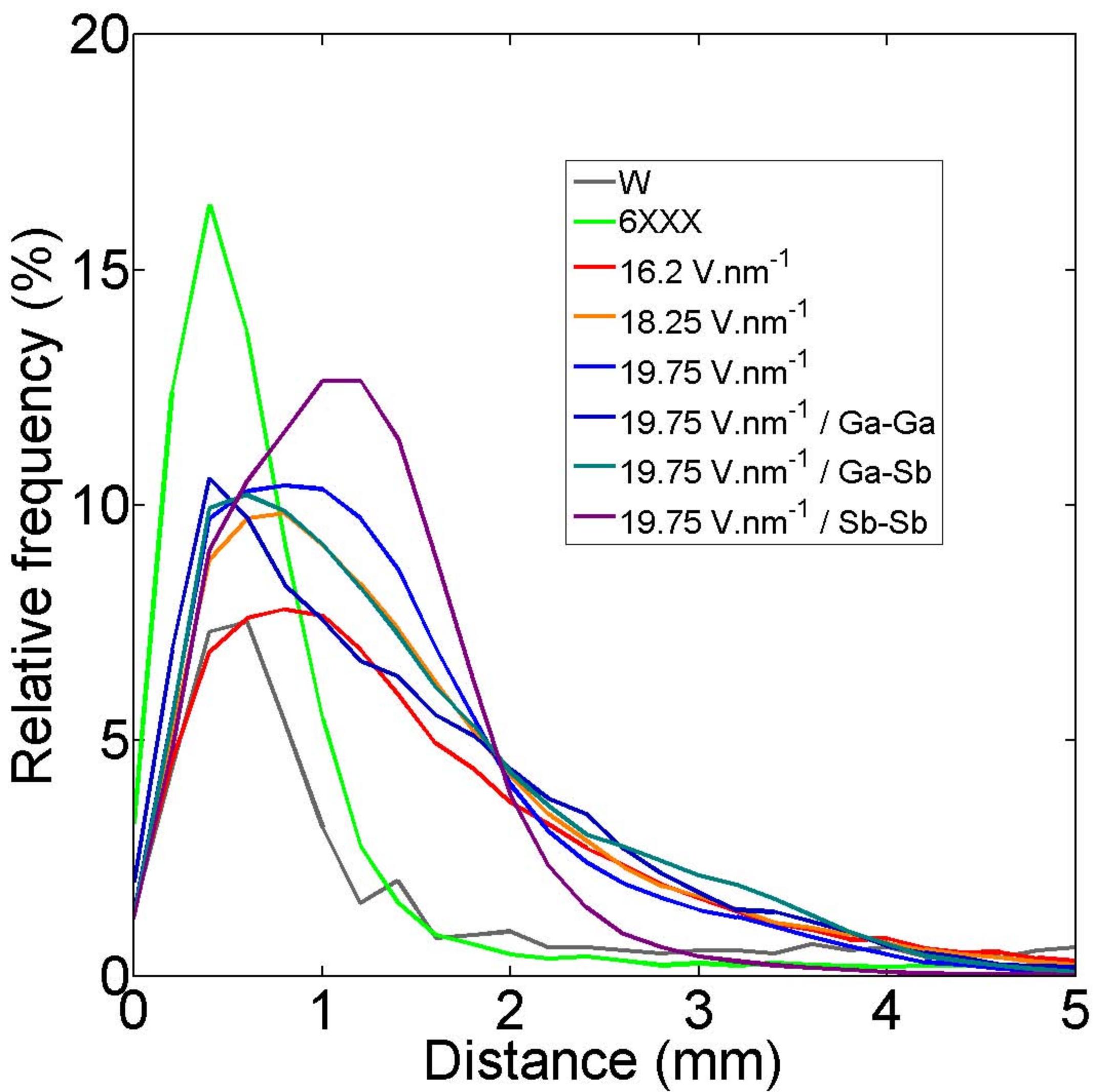

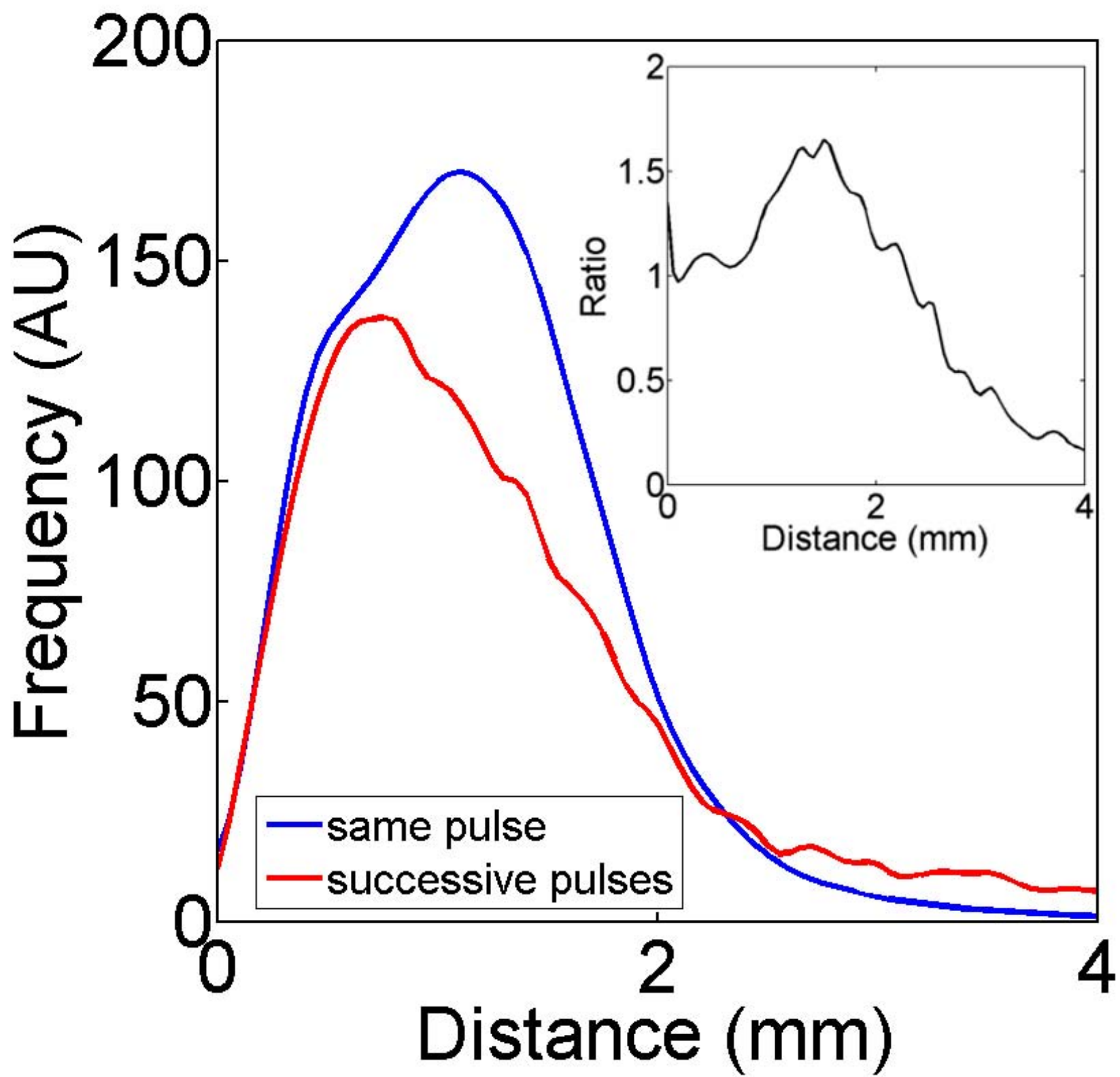

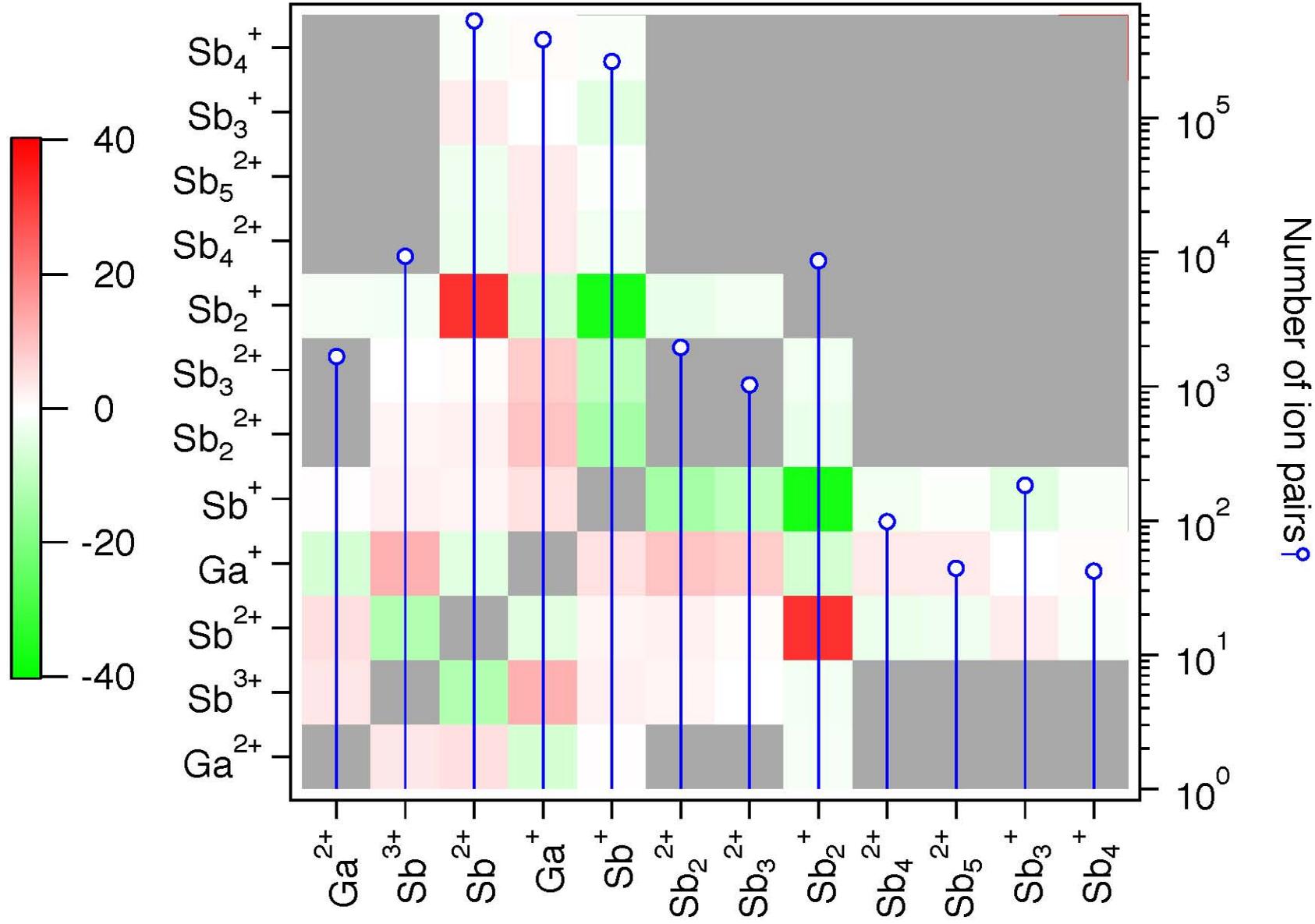